# Protection of the Rights of Large Families as One of the Key Tasks of the State's Social Policy

**Valery Dolgov,**
*Associate Professor, Russian University of Transport,*
*Moscow, Russia*

**Mattia Masolletti,**
*Associate Professor, NUST University*
*Rome, Italy*

**Abstract**

The authors of the article analyze the policy of the Russian government in the field of family support, paying attention to legal programs at the federal and regional levels. The maternity capital program is considered separately, as well as measures aimed at supporting large families.

## 1. Introduction

In the Russian Federation, the family, motherhood, fatherhood and childhood are protected by the state in accordance with Article 38 of the Constitution of the Russian Federation. The State family policy is an integral part of the social policy of the Russian Federation and represents an integral system of principles, assessments and measures of an organizational, economic, legal, scientific, informational, propaganda and personnel nature aimed at improving the conditions and improving the quality of family life.

Most of the problems of the Russian society in one way or another leave their mark on the normal functioning of a young family, creating certain difficulties for it. And since a young family is at the stage of formation and development, it needs special, comprehensive support and requires increasing the role of the state in creating normal conditions for achieving a level of well-being.

## 2. Main part

Currently, there are four main forms of social protection of families with children in Russia:

1. Monetary payments to the family for children in connection with the birth, maintenance and upbringing of children (allowances and pensions).

2. Labor, tax, housing, credit, medical and other benefits for families with children, parents and children.

3. Legal, medical, psychological, pedagogical and economic counseling, parental general education, scientific and practical conferences and congresses.

4. Free distribution of baby food, medicines, clothing and shoes, food for pregnant women to families and children.

5. Federal, regional, targeted social programs such as 'Family Planning', 'Children of Russia'.

The improvement of the relevant regulatory framework is of great importance for the sphere of social protection. The following laws have been developed and adopted: 'On state benefits for citizens with children', 'On the procedure for assigning and paying monthly compensation to women with children under the age of 3 who were dismissed in connection with the liquidation of enterprises, institutions, organizations', 'On improving the system of state social benefits and compensation payments to families with children and increasing their amounts', 'On compensation payments to families with children, students and other categories of persons', 'On the state system for the prevention of neglect and juvenile delinquency, protection of their rights'. To solve specific tasks of social protection of the family, the program-target method has become more widely used. In particular, the Federal Program 'Children of Russia' had been developed and adopted for implementation, which includes 6 target programs: 'Disabled Children', 'Orphans', 'Children of the North', 'Development of the Baby Food Industry', 'Family Planning'.

The legislation of the Russian Federation in the field of social policy defines the main directions of ensuring the rights of citizens of the Russian Federation, enshrined in the Constitution. In this regard, the current Russian legislation in the field of social policy can be conditionally classified, taking the rights of citizens guaranteed by Articles 37-44 of the Constitution of the Russian Federation in the sphere of [7]:

- labor relations and the rights of citizens to rest (the Labor Code of the Russian Federation, the Law of the Russian Federation (adopted on April 19, 1991) No. 1032-I 'On employment of the population in the Russian Federation', the Federal Law (adopted on February 23, 1995) No. 26-FZ 'On natural healing resources, health-improving areas and resorts', the Federal Law (adopted on October 34, 1997) No. 134-FZ 'On the subsistence minimum', etc.);

- protection of motherhood and childhood (the Federal Law No. 81-FZ, adopted on May 19, 1995 – 'On state benefits to citizens with children'; the Federal Law No. 159-FZ, adopted on December 21, 1996 – 'On additional guarantees for the support of orphans and children left without parental care'; the Federal Law No. 124-FZ, adopted on July 24, 1998 – 'On basic guarantees of the rights of the child in the Russian Federation', etc.);

- social security (the Federal Law No. 195-FZ, adopted on December 10, 1995 – 'On the basics of social services for the population in the Russian Federation'; the Federal Law No. 181-FZ, adopted on November 24, 1995 – 'On Social Protection of Disabled People'; the Federal Law No. 178-FZ, adopted on July 17, 1999 – 'On State Social Assistance', etc.);

- housing policy (the Housing Code of the Russian Federation, the Law of the Russian Federation No. 5242-1, adopted on June 25, 1993 – 'On the right of citizens of the Russian Federation to freedom of movement, choice of place of stay and residence within the Russian Federation'; the Law of the Russian Federation No. 1541-1, adopted on July 4, 1991, - 'On Privatization of Housing Stock in the Russian Federation'; the Federal Law No. 102-FZ, adopted on July 16, 1998, - 'On Mortgage (pledge of real estate)', etc.);

- health (the Law of the Russian Federation, adopted on June 28, 1991, No. 1499-1 'On medical insurance of citizens of the Russian Federation'; the Law of the Russian Federation, adopted on June 9, 1993, No. 5142-I 'On the donation of blood and its components'; the Law of the Russian Federation No. 5487-1 'On the fundamentals of the legislation of the Russian Federation on health protection of citizens', adopted on 22 July 1993, etc.);

- the environment (the law of the Russian Federation, adopted on May 15, 1991, № 1244-1 'On social protection of citizens exposed to radiation as a result of the Chernobyl accident'; the Federal Law, adopted on March 30, 1999, No. 52-FZ – 'On the sanitary and epidemiological welfare of the population', etc.);

- education (Law of the Russian Federation, adopted on July 10, 1992, No. 3266-1 'On Education'; the Federal Law, adopted on August 22, 1996, No. 125-FZ 'On Higher and Postgraduate Education', etc.);

- culture (the Law of the Russian Federation No. 3612-I, adopted on October 9, 1992, 'the Fundamentals of the Legislation of the Russian Federation on culture'; the Federal Law No. 54-FZ, adopted on May 26, 1996, 'On the Museum Fund of the Russian Federation and Museums in the Russian Federation'; the Federal Law No. 126-FZ, adopted on August 22, 1996, 'On State Support of Cinematography of the Russian Federation', etc.).

Article 114 of the Constitution of the Russian Federation establishes the powers of the Government of the Russian Federation to conduct a unified state policy in the field of culture, science, education, healthcare, social security, and ecology in the country. The main direction of the policy in the social sphere is to take care of a person, create conditions for his or her decent life and comprehensive development.

In order to implement the provisions of the Constitution of the Russian Federation guaranteeing the social rights of citizens of the Russian Federation, the Federal Assembly of the Russian Federation has developed and adopted more than 30 Federal laws, some of which have already been mentioned above. The responsible federal executive authorities, by executing the secondary legislation of the President of the Russian Federation (Decrees and Orders) and the Government of the Russian Federation (resolutions and orders) aimed at implementing the current Russian legislation in the social sphere, as well as their own acts of federal executive authorities in this area, implement the legal foundations of social policy at the federal level.

The regional level determines the directions and mechanisms for implementing social policy in each specific region (entity) of the Russian Federation by developing and adopting acts of legislative (representative) and executive authorities of the regions of the Russian Federation.

The main powers of the state authorities of the regions of the Russian Federation in the field of social policy are defined by the Constitution of the Russian Federation and the Federal Law No. 184-FZ (dated October 6, 1999) 'On the general principles of the organization of legislative (representative) and executive authorities of the regions of the Russian Federation'.

The local level is made up of municipal acts. The main powers of local self-government bodies of municipalities in the field of social policy are defined by the Constitution of the Russian Federation and the Federal Law No. 131-FZ (dated October 6, 2003) 'On General Principles of the Organization of Local Self-Government in the Russian Federation'.

The purpose of the state policy in relation to the modern young family is the formation and development of a prosperous young family and improving the quality of its life, ensuring that the young family performs socio-demographic functions, including stimulating the birth rate of children and their upbringing. The objectives of the state youth family policy are:

- legislative provision of an independent status of an object of state family policy to a young family and the practical realization of the potential of this status in all spheres of its life;

- ensuring that the state respects the rights of a young family in solving social problems;

- improving the system of state social guarantees to ensure the achievement of the level of well-being of young families;

- strengthening the institution of the Russian family on the basis of traditional folk socio-cultural values, spirituality and national way of life;

- formation of a positive 'pro-family' public opinion, promotion of a family lifestyle, increasing the prestige of a socially prosperous family [1];

- ensuring the preservation of the family environment as an environment for personal self-development and self-realization of spouses, reproduction, upbringing and development of children-full-fledged citizens of the Russian society;

- taking into account the interests of a young family in the process of spiritual, moral, economic and socio-cultural development of society;

- providing young families with the necessary information support in their formation and stable life;

- development and support of public organizations of young families.

In particular, social protection of the family is a multi-level system of mainly state measures to ensure minimum social guarantees, rights, benefits and freedoms of a normally functioning family in a situation of risk in the interests of the harmonious development of the family, the individual and society.

Nowadays in the Russian Federation state measures of the family support have four main forms of implementation [2]:

- monetary - in the form of compensation and lump-sum payments in connection with the birth and upbringing of children, including the program of maternal (family) capital; provision of monthly cash payments for the third child; subsidies for housing and utilities;

- services - by organizing recreation and improving the health of children; by providing social services for families with children, neglected and street children (providing specific psychological, legal, pedagogical assistance, counseling, etc.);

- natural form – by providing families with housing at the expense of budgetary funds; providing free travel on public transport, and provision of fuel; free issue the family with kids (baby food, food for pregnant women clothing and shoes, medical supplies, etc.);

- various benefits for the family with children, parents or children (labor, tax, housing, credit, medical, and others, including the free provision of land for individual housing construction), the list of which depends on the capabilities of the RF subject.

Then we would like to elaborate on the 'Maternity Capital' program in more detail (see *Fig.1*). This program was launched on January 1, 2007 [6]. More than 7.3 million families, in which a second or subsequent child was born or adopted during this period, received the state certificate of MSK for 10 years. For 10 years, the size of the certificate has increased from 250,000 rubles to 453,026 rubles. The program has been extended at the present time.

Since 2018, families have been given more opportunities to use the maternity capital's certificate immediately after the birth or adoption of a second child:

1. Monthly payment from the maternity capital for a family with low incomes (less than 1.5 of the subsistence minimum of the working-age population per person in the family) in case of the birth of a second child from January 1, 2018.

2. Preschool education, child care and supervision. Starting in 2018, families receive financial support for pre-school education almost immediately after the birth of a child, since now the maternity capital can be used already two months after the acquisition of the right to it. Previously, it was possible to use money for these purposes only three years after the birth or adoption of a child for whom maternity capital was issued. You can use the funds in such a period to pay for a kindergarten and a nursery, including private ones, as well as for the payment of services for the care and supervision of a child. In both cases, a necessary condition is that the organization has a license to provide the relevant services.

3. Preferential mortgages for families with two and three children Russian families who will have a second or third child in 2018-2021 will be able to take advantage of preferential credit conditions to improve their living conditions. Preferential mortgages can also be repaid with maternity capital. It is not necessary to wait for the three-year-old child who has given the right to a certificate.

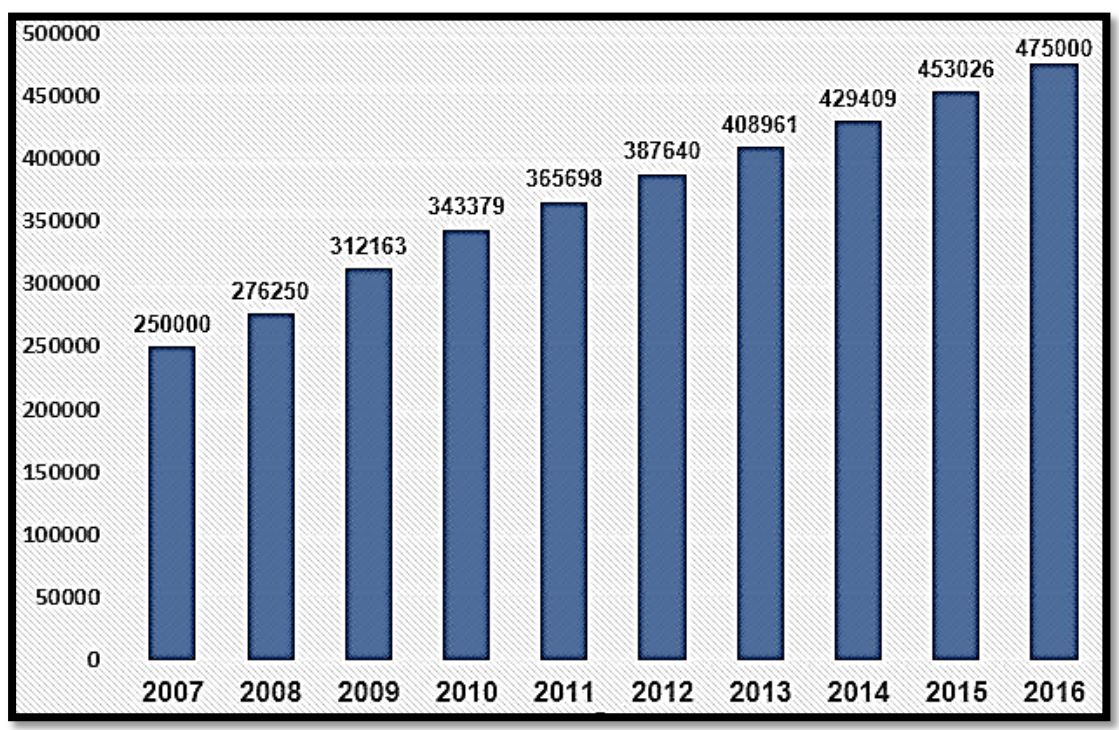

Fig.1. Indexation of maternity capital in Russia by year (in rubles)
*Source: the Ministry of Economic Development of the Russian Federation, 2007-2016*

In addition to the maternity capital provided at the expense of the federal budget, 72 constituent entities of the Russian Federation have introduced regional maternity capital financed by the budget of the constituent entities of the Russian Federation at the birth of the third (subsequent) child. In addition, in accordance with paragraph 2 of the Decree of the President of the Russian Federation No. 606, adopted on May 7, 2012, 'On measures to implement the demographic policy of the Russian Federation', the Russian regions are recommended to establish a monthly cash payment to families in need of support in the amount of the subsistence minimum for children determined in the region, assigned in case of birth after December 31, 2012 of the third child or subsequent children, until the child reaches the age of three years. This monthly cash payment is established in 69 regions of the country.

The most extensive form of state support for the family is a complex of monetary payments: social benefits and compensation payments, which are addressed to parents, a family with children. A unified system of state benefits for families with children in connection with their birth and upbringing, providing guaranteed state material support for motherhood, fatherhood and childhood, which are established by the Federal Law 'On State Benefits for Citizens with children', which defines the rights of a family of children to receive benefits related to federal benefits, i.e. those whose appointment rules are the same throughout the entire territory of Russia. At the same time, each of these benefits, establishing the right to receive and the required amount of specific types of financial assistance, has its own specifics in terms of the rules and conditions of appointment, the range of recipients, etc.

In our opinion, the most important measure in the field of state support for the family was the law signed by the President of the Russian Federation on March 7, 2018 on increasing the minimum wage (from 9 thousand up to 11 thousand rubles; since 2021 it was increased to 12 thousand 792 rubles), equating this amount with the subsistence minimum, which eventually leads to recalculation of the amount of child benefits and sick leave (related to the illness of a child or parent).

In addition to federal cash payments and benefits, many regions of the Russian Federation provide their own regional benefits to families with children. In contrast to national payments, the procedure for granting and the amount of regional, regional, and republican benefits are regulated by regulatory legal acts of local authorities. In this regard, in different regions, not only different amounts of such assistance are established, but also different types and grounds for receiving it [4]. This largely depends on the level of socio-economic development of a particular region and the capabilities of the local budget. Thus, the right to receive child benefits, which is paid until the child turns 16, refers to regional laws on monthly child benefits, which in each subject of the federation is paid in a certain amount by the subject and according to its own rules.

The regions of Russia have the right to introduce other benefits that are not provided for by federal legislation: an allowance for the birth of a child in a large family, an allowance for unemployed pregnant women registered with the employment service, compensation for not providing a place in kindergarten, and so on. For instance, the Law of the Krasnoyarsk Territory 'On additional measures to support families with children in the Krasnoyarsk Territory', 'On monthly child allowance' [5].

At the regional level, support is also provided to large families, providing them with guarantees, benefits and compensations aimed at social support for improving the living conditions of large families, improving their status and improving the situation of children in them, which are not provided for by federal legislation. The main benefits for large families in all subjects of the Russian Federation include, for example, the admission of children to preschool institutions in the first place, free meals (breakfast and lunch) for students in educational institutions. Local authorities also have the right to establish additional benefits for this category of citizens living in the territory under their jurisdiction. These benefits depend entirely on their budget and on the conditions set by the legislators of a particular region for receiving them. Among them are the Krasnoyarsk Territory, the Republic of Buryatia, Arkhangelsk, Amur and other regions. Thus, citizens with three or more children have the right to provide a land plot provided by a local self-government body authorized to manage and dispose of (land plots), on the basis of a written application for the free provision of a land plot. In addition, in many regions, utility tariffs have been reduced for large families. For those who live in houses where there is no central heating, there are preferential prices for coal and other types of fuel.

### 3. Conclusion

Thus, at the present stage, state support for the family in Russia faces the following key contradictions that require a speedy settlement:

- The essence of the problem of implementing family policy is the discrepancy between the provisions of the Concept of State Family Policy in the Russian Federation for the period up to 2025 and the actual state of the Russian family in the conditions of an escalating socio-economic crisis.
- There are gaps in the activities of executive bodies designed to implement state policy in the field of family and marriage, as well as in tasks and functions of its implementation:

a) 'drown' among many other tasks assigned to these bodies [3];

b) narrowed down to the protection of the rights of children from disadvantaged families.

### References

[1] Barotskaya K.B. (2011) The right of the family to social protection // Social and pension law. - No. 3. P. 6.

[2] Sidorina T.Yu. (2010) Two centuries of social policy. - Moscow: Publishing House of the Russian State University.